\documentstyle[12pt]{article}
\huge
\renewcommand{\theequation}{\arabic{section}.\arabic{equation}} 
\def\setzero{\setcounter{equation}{0}}

\newcounter{eqalph}
\def\bph{\setcounter{eqalph}{\value{equation}}
   \addtocounter{eqalph}{1}
   \setcounter{equation}{0}
   \renewcommand{\theequation}{\arabic{section}.\arabic{eqalph}\alph{equation}}}
\def\eph{\setcounter{equation}{\value{eqalph}}
   \renewcommand{\theequation}{\arabic{section}.\arabic{equation}}
\par\noindent}

\begin{document}

\baselineskip 18pt
\newcommand{\ii}{\mbox{i}}
\def \sech{{\rm sech}}
\def \tanh{{\rm tanh}}
\def \cn{{\rm cn}}
\def \sn{{\rm sn}}
\def\bm#1{\mbox{\boldmath $#1$}}
\newfont{\bg}{cmr10 scaled\magstep4}
\newcommand{\bzr}{\smash{\hbox{\bg 0}}}
\newcommand{\bzl}{%
   \smash{\lower1.7ex\hbox{\bg 0}}}
\title{Hungry Volterra equation, 
multi boson  KP hierarchy
\\
and \\
Two Matrix Models } 
\date{\today}
\author{ Masato {\sc Hisakado}  
\\
\bigskip
\\
{\small\it Department of Pure and Applied Sciences,}\\
{\small\it University of Tokyo,}\\
{\small\it 3-8-1 Komaba, Meguro-ku, Tokyo, 153, Japan}}
\maketitle

\vspace{20 mm}

Abstract

We consider the hungry Volterra hierarchy 
from the view point of 
the multi boson  KP hierarchy.
We construct  the hungry Volterra equation as the 
B\"{a}cklund transformations (BT) which are not the 
ordinary  ones.
We call them ``fractional '' BT.
We also study the relations 
between the (discrete time) hungry Volterra equation and 
two matrix models.
{}From this point of view we study  
the reduction from (discrete time ) 2d Toda lattice to the 
(discrete time ) hungry Volterra equation.


\vfill
\par\noindent
{\bf  }

\newpage

\section{Introduction}
The $W$ algebra 
was introduced by Zamolodchikov
as an additional symmetry of the conformal field theory.\cite{za}
The classical $W_{N}$ algebra can be  constructed  from the free
fields $r_{i}$ for $i=1,2,\cdots,N$ satisfying the Poisson 
algebra
\begin{equation}
\{r_{i}(x),r_{j}(y)\}
=(-\delta_{i,j}+\frac{1}{N})
\delta(x-y).
\end{equation}
The Poisson structure is related $N$-reduced 
KP hierarchy.
The Poisson map from the free fields to the pseudo differential 
 operator $L$ is defined 
as

\begin{equation}
L=(\partial +r_{1}(x))(\partial+r_{2}(x))\cdots(\partial+r_{N}(x))
\end{equation}
with $\partial\equiv\partial /\partial x$, and 
the constraint
\begin{equation}
\sum_{i=1}^{N}r_{i}(x)=0.
\end{equation}

Recently  deformations of the $W$ algebra received much attention.
One is the $q$ deformation of the $W$ algebra.
Another  deformation of  the $W$ algebra 
is lattice $W$ algebra,
which may be related to the $sl_{N}$ Toda theory.\cite{fa}
The most famous example is the Lotka-Volterra model.
It is well known that the Lotka-Volterra model reduces to the 
 Korteweg-de Vries  (KdV) equation in the continuous limit, and that 
it can be formulated in terms of lattice Virasoro
 algebra.

In this paper we shall study  
the hungry Volterra equation (or `` Bogoyavlensky lattice'')
which is known as an extended Volterra model.\cite{Bo}
It  
is pointed out that the hungry Volterra equation  is a fundamental 
integrable system of the lattice $W_{N}$ algebra.\cite{hi}
We define the $k$-hungry Volterra equation as
\begin{equation}
\frac{{\rm d}V_{n}}{{\rm d}t_{1}}
=V_{n}
(\sum_{i=1}^{k-1}V_{n+i}-\sum_{i=1}^{k-1}V_{n-i}).
\label{BL}
\end{equation}
In the case $k=1$ 
(\ref{BL}) becomes the Lotka-Volterra equation.

We consider the hungry Volterra equation 
as B\"{a}cklund transformations (BT) of the multi boson  KP equation 
which is related 
to the $sl(k+1, k)$ algebra.\cite{A1}-\cite{box}
We  call the BT 
``fractional BT''.
If we repeat  the fractional BT  $k$ times 
we can obtain the usual BT.
To study  relations 
between the 2d Toda lattice and the hungry Volterra equation
we consider  the hungry Volterra equation
in  two matrix models.
The  most simple model becomes  a matrix model for 
the bi-colored random triangulation.
{}Furthermore we 
consider the discrete time hungry Volterra equation
in   two matrix models  with 
 Penner type potential.
It becomes a simple relation of the partition functions 
in  bilinear form.

This paper is organized as follows.
In  section 2 	
we consider the eigenvalue problems of the 3-hungry Volterra equation.
In section 3 
we define the fractional BT and 
 it is equivalent to  the $k$-hungry Volterra equation.
In section 4 
we consider the reduction 
from the 3-hungry Volterra equation 
 to the classical $W_{3}$ and Virasoro algebra.
In  section 5
 we obtain the hungry Volterra equation  and the discrete time 
hungry Volterra equation 
in  two matrix models.
The last section is devoted to the concluding remarks.

\setzero
\section{Hungry Volterra equation}

We consider the spectral problems,
\begin{eqnarray}
\lambda^{1/3}\hat{\Psi}_{n+\frac{2}{3}}
&=&\Psi_{n+1}+u_{n}\Psi_{n},
\nonumber \\
\lambda^{1/3}\tilde{\Psi}_{n+\frac{1}{3}}
&=&\hat{\Psi}_{n+\frac{2}{3}}+v_{n}\hat{\Psi}_{n-\frac{1}{3}},
\nonumber \\
\lambda^{1/3}\Psi_{n}
&=&\hat{\Psi}_{n+\frac{1}{3}}+w_{n}\hat{\Psi}_{n-\frac{2}{3}}.
\label{see}
\end{eqnarray} 
From (\ref{see})
we can obtain
\begin{eqnarray}
\lambda\Psi_{n}&=&\Psi_{n+1}+a_{0}(n)\Psi_{n}
+a_{1}(n)\Psi_{n-1}
+a_{2}(n)\Psi_{n-2},
\nonumber \\
\lambda\tilde{\Psi}_{n+\frac{1}{3}}&=&
\tilde{\Psi}_{n+\frac{4}{3}}+\tilde{a}_{0}(n)
\tilde{\Psi}_{n+\frac{1}{3}}
+\tilde{a}_{1}(n)\tilde{\Psi}_{n-\frac{2}{3}}
+\tilde{a}_{2}(n)\tilde{\Psi}_{n-\frac{5}{3}},
\nonumber \\
\lambda\hat{\Psi}_{n+\frac{2}{3}}&=&
\hat{\Psi}_{n+\frac{5}{3}}+\hat{a}_{0}(n)
\hat{\Psi}_{n+\frac{2}{3}}
+\hat{a}_{1}(n)\hat{\Psi}_{n-\frac{1}{3}}
+\hat{a}_{2}(n)\hat{\Psi}_{n-\frac{4}{3}},
\label{toda}
\end{eqnarray}
where 
\begin{eqnarray}
a_{0}(n)&=&u_{n}+v_{n}+w_{n},\;\;\;
a_{1}(n)=v_{n}u_{n-1}+u_{n-1}w_{n}+w_{n}v_{n-1},\;\;\;
a_{2}(n)=w_{n}v_{n-1}u_{n-2},
\nonumber \\
\tilde{a}_{0}(n)&=&w_{n+1}+u_{n}+v_{n},\;\;\;
\tilde{a}_{1}(n)=u_{n}w_{n}+w_{n}v_{n}+v_{n}u_{n-1},\;\;\;
\tilde{a}_{2}(n)=v_{n}u_{n-1}w_{n-1},
\nonumber \\
\hat{a}_{0}(n)&=&v_{n+1}+w_{n+1}+u_{n},\;\;\;
\hat{a}_{1}(n)=w_{n+1}v_{n}+v_{n}u_{n}+u_{n}w_{n},\;\;\;
\hat{a}_{2}(n)=u_{n}w_{n}v_{n-1}.
\end{eqnarray}
Using (\ref{toda}) we can obtain the
first lattice flow $\partial=\partial_{x}$
\begin{eqnarray}
\partial \Psi_{n}&=&\Psi_{n+1}+a_{0}(n)\Psi_{n},
\nonumber \\
\partial\tilde{\Psi}_{n+\frac{1}{3}}&=&
\tilde{\Psi}_{n+\frac{4}{3}}+\tilde{a}_{0}(n)
\tilde{\Psi}_{n+\frac{1}{3}},
\nonumber \\
\partial 
\hat{\Psi}_{n+\frac{2}{3}}&=&
\tilde{\Psi}_{n+\frac{5}{3}}+\hat{a}_{0}(n)
\tilde{\Psi}_{n+\frac{2}{3}}.
\label{flow}
\end{eqnarray}
{}From (\ref{toda}) and (\ref{flow}) we can obtain the consistency conditions,
\begin{eqnarray}
\frac{{\rm d} a_{0}(n)}{{\rm d}x}&=&a_{1}(n+1)-a_{1}(n),
\nonumber \\
\frac{{\rm d} a_{1}(n)}{{\rm d}x}&=&a_{1}(n-1)+\frac{{\rm d} a_{0}(n-1)}
{{\rm d}x},
\nonumber \\
\frac{{\rm d} a_{2}(n)}{{\rm d}x}&=&a_{2}(a_{0}(n)-a_{0}(n-2)).
\end{eqnarray}
In the same way we can obtain same equations 
about $\tilde{a}(n)$ and $\hat{a}(n)$.
These are the generalized Toda equations.

On the other hands, using the consistency condition of (\ref{see})
and (\ref{flow})
the equations of motion
read
\begin{eqnarray}
\frac{{\rm d} u_{n}}{{\rm d} x}&=&u_{n}(-v_{n}-w_{n}+v_{n+1}+w_{n+1}),
\nonumber \\
\frac{{\rm d} v_{n}}{{\rm d} x}&=&v_{n}(-u_{n-1}-w_{n}+u_{n}+w_{n+1}),
\nonumber \\
\frac{{\rm d} w_{n}}{{\rm d} x}&=&w_{n}(-u_{n-1}-v_{n-1}+u_{n}+v_{n}).
\label{em}
\end{eqnarray}
These are  3-hungry Volterra equations.
(\ref{see}) and (\ref{flow}) can be cast into the form
\begin{equation}
\lambda\Psi_{n}=L_{n}\Psi_{n},
\;\;\;
\lambda\tilde{\Psi}_{n+\frac{1}{3}}=
\tilde{L}_{n+\frac{1}{3}}\tilde{\Psi}_{n+\frac{1}{3}},
\;\;\;
\lambda\hat{\Psi}_{n+\frac{2}{3}}=
\hat{L}_{n+\frac{2}{3}}\hat{\Psi}_{n+\frac{2}{3}},
\end{equation}
where 
\begin{eqnarray}
L_{n}&=&(\partial -v_{n-1}-u_{n-1})(\partial -v_{n-1}-u_{n-1}-w_{n})^{-1}
(\partial -u_{n-1}-w_{n})
\nonumber \\
& &
(\partial-u_{n-1}-w_{n}-v_{n})^{-1}
(\partial -w_{n}-v_{n}),
\nonumber \\
\tilde{L}_{n+\frac{1}{3}}&=&
(\partial -u_{n-1}-w_{n})(\partial-u_{n-1}-w_{n}-v_{n})^{-1}
(\partial -w_{n}-v_{n})
\nonumber \\
& &
(\partial -v_{n}-w_{n}-u_{n})^{-1}
(\partial -u_{n}-v_{n}),
\nonumber \\
\hat{L}_{n+\frac{2}{3}}&=&
(\partial -w_{n}-v_{n})(\partial -v_{n}-w_{n}-u_{n})^{-1}
(\partial -u_{n}-v_{n})
\nonumber \\
& &
(\partial -u_{n}-v_{n}-w_{n+1})^{-1}
(\partial -u_{n}-w_{n+1}).
\label{lax}
\end{eqnarray}
This is the form of the 4-boson KP hierarchy.
Using (\ref{toda}) and (\ref{flow})
we can obtain
the B\"{a}cklund Transformations
\begin{eqnarray}
L_{n+1}&=&(\partial -a_{0}(n))L_{n+1}(\partial -a_{0})^{-1},
\nonumber \\
\tilde{L}_{n+1+\frac{1}{3}}&=&
(\partial -\tilde{a}_{0}(n))\tilde{L}_{n+\frac{1}{3}}
(\partial -\tilde{a}_{0}(n))^{-1},
\nonumber \\
\hat{L}_{n+1+\frac{2}{3}}&=&
(\partial -\hat{a}_{0}(n))\hat{L}_{n+\frac{1}{3}}
(\partial -\hat{a}_{0}(n))^{-1}.
\label{bt1}
\end{eqnarray}

\setzero
\section{Fractional B\"{a}cklund Transformations}

We consider multi boson KP hierarchy:
\begin{equation}
L^{(k)}=(\partial -q_{1})(\partial-\tilde{q}_{1})^{-1}
(\partial -q_{2})
(\partial-\tilde{q}_{2})^{-1}
\cdots
(\partial-\tilde{q}_{k-1})^{-1}(\partial -q_{k}),
\end{equation}
with
a Dirac constraint
\begin{equation}
\sum_{j=1}^{k}q_{j}-\sum_{l=1}^{k-1}\tilde{q}_{l}=0.
\label{dc}
\end{equation}

We construct ``fractional'' B\"{a}cklund transformations (BT),
\begin{equation}
L^{(k)}_{n+\frac{l+1}{k}}
=
(T_{n+\frac{l}{k}}^{(k)})^{-1}L^{(k)}_{n+\frac{l}{k}}T^{(k)}_{n+\frac{l}{k}},
\label{fbt}
\end{equation}
where
\begin{equation}
T_{n+\frac{l}{k}}^{(k)}=
(\partial -q_{1}^{(n+\frac{l}{k})})(\partial-\tilde{q}_{1}^{(
n+\frac{l}{k})})^{-1}.
\end{equation}
Here we realize that the normal lattice jump
$n\rightarrow n+1$ can be given a meaning of  the
BT,
\begin{eqnarray}
L^{(k)}_{n+1+\frac{l}{k}}
&=&(\partial -q_{1}^{(n+1+\frac{l}{k})})(\partial-\tilde{q}_{1}^{(n+1+\frac{l}{k})})^{-1}
(\partial -q_{2}^{(n+1+\frac{l}{k})})
(\partial-\tilde{q}_{2}^{(n+1+\frac{l}{k})})^{-1}
\cdots
\nonumber \\
&\times&
(\partial-\tilde{q}_{k-1}^{(n+1+\frac{l}{k})})^{-1}(\partial -q_{k}^{(n+1+\frac{l}{k})}),
\end{eqnarray}
and 
\begin{equation}
L^{(k)}_{n+1+\frac{l}{k}}=
({\cal T}_{n+\frac{l}{k}}^{(k)})^{-1}L^{(k)}_{n+\frac{l}{k}}{\cal T}^{(k)}_{n+\frac{l}{k}}.
\label{bt}
\end{equation}
Using (\ref{fbt}) and (\ref{bt})
we can obtain a  relation
\begin{equation}
{\cal T}_{n+\frac{l}{k}}^{(k)}=
T_{n+1+\frac{l}{k}}^{(k)}
T_{n+\frac{l+k-1}{k}}^{(k)}
\cdots
T_{n+\frac{l}{k}}^{(k)}.
\label{fab}
\end{equation}
{}From  (\ref{fbt})
we get 
\begin{eqnarray}
T_{n+\frac{l}{k}}^{(k)}&=&
(\partial -q_{1}^{(n+\frac{l}{k})})(\partial-\tilde{q}_{1}^
{(n+\frac{l}{k})})^{-1}
\nonumber \\
&=&
(\partial-\tilde{q}_{k-1}^
{(n+\frac{l+1}{k})})^{-1}
(\partial -q_{k}^{(n+\frac{l+1}{k})})
=
(\partial-\tilde{q}_{k-1}^
{(n+\frac{l+1}{k})})^{-1}
(\partial -q_{1}^{(n+1+\frac{l}{k})}).
\label{ec}
\end{eqnarray}
To obtain the last equation
we use  relations
\begin{equation}
(\partial -q_{k}^{(n+\frac{l+1}{k})})
=
(\partial -q_{k-1}^{(n+\frac{l+2}{k})})
=\cdots
=
(\partial -q_{1}^{(n+1+\frac{l}{k})}).
\end{equation}
To satisfy (\ref{dc})  using (\ref{ec})
we can set
\begin{equation}
- q_{1}^{(n+\frac{l}{k})}+\tilde{q}_{1}^{(n+\frac{l}{k})}
=
\tilde{q}_{k-1}^
{(n+\frac{l+1}{k})}-q_{1}^{(n+1+\frac{l}{k})}\equiv p_{n+\frac{l}{k}}.
\label{dc1}
\end{equation}
Using (\ref{dc1}) we can rewrite (\ref{ec})
\begin{equation}
(\partial  -q_{1}^{(n+\frac{l}{k})})
(\partial -q_{1}^{(n+\frac{l}{k})}-p_{n+\frac{l}{k}})^{-1}
=
(\partial -q_{1}^{(n+1+\frac{l}{k})}-p_{n+\frac{l}{k}})^{-1}
(\partial -q_{1}^{(n+1+\frac{l}{k})}).
\label{hvvv}
\end{equation}
Then (\ref{hvvv}) becomes  an equation 
\begin{equation}
\frac{{\rm d}p_{n+\frac{l}{k}}}{{\rm d}x}=p_{n+\frac{l}{k}}
(q_{1}^{(n+1+\frac{l}{k})}-q_{1}^{(n+\frac{l}{k})}).
\label{hvk}
\end{equation}
Using the  fractional BT recursively we can obtain
\begin{eqnarray}
L_{n+1+\frac{l}{k}}&=&
(\partial -q_{1}^{(n+1+\frac{l}{k})})
(\partial -q_{1}^{(n+1+\frac{l}{k})}-p_{n+1+\frac{l}{k}})^{-1}
(\partial -q_{2}^{(n+1+\frac{l}{k})})
(\partial -q_{2}^{(n+1+\frac{l}{k})}-p_{n+\frac{l+k-1}{k}})^{-1}
\cdots
\nonumber \\
&\times&
(\partial -q_{k-1}^{(n+1+\frac{l}{k})}-p_{n+\frac{l+1}{k}})^{-1}
(\partial -q_{k}^{(n+1+\frac{l}{k})}).
\end{eqnarray}
{}From (\ref{dc}) we get
\begin{equation}
q_{k}^{(n+1+\frac{l}{k})}=\sum_{j=1}^{k-1}p_{n+\frac{l+j}{k}}.
\label{jj}
\end{equation}
If we substitute (\ref{jj}) into (\ref{hvk}),
we  can obtain 
\begin{equation}
\partial p_{n+\frac{l}{k}}=p_{n+\frac{l}{k}}
(\sum_{j=1}^{k-1}p_{n+\frac{l+j}{k}}-\sum_{j=1}^{k-1}p_{n-1+\frac{l+j}{k}}).
\label{hvk1}
\end{equation}
(\ref{hvk1}) ia nothing but the 
$k$-hungry Volterra equation.
Using (\ref{fab}) we can obtain the 
normal lattice BT
\begin{equation}
{\cal T}_{n+\frac{l}{k}}^{(k)}
=
(\partial -
\sum_{j=1}^{k}p_{n+\frac{l+j}{k}}).
\end{equation}
In the case $k=3$
it corresponds to (\ref{bt1}).

\setzero
\section{$W$ algebra and  Torus}

We can rewrite the Lax operator (\ref{lax})
using (\ref{em}) or directly from (\ref{flow}),
\begin{eqnarray}
L_{n}&=&(\partial -u_{n}-v_{n}-w_{n})^{-1}
(\partial -u_{n+1}-v_{n+1}-w_{n+1})^{-1}
\nonumber \\
& &
(\partial -u_{n+1}-v_{n+1})(\partial -u_{n}-w_{n+1})
(\partial -v_{n}-w_{n}),
\nonumber \\
\tilde{L}_{n+\frac{1}{3}}&=&
(\partial -u_{n}-v_{n}-w_{n+1})^{-1}
(\partial -u_{n+1}-v_{n+1}-w_{n+2})^{-1}
\nonumber \\
& &
(\partial -u_{n+1}-w_{n+2})(\partial -v_{n+1}-w_{n+1})
(\partial -u_{n}-v_{n}),
\nonumber \\
\hat{L}_{n+\frac{2}{3}}&=&
(\partial -u_{n}-v_{n+1}-w_{n+1})^{-1}
(\partial -u_{n+1}-v_{n+2}-w_{n+2})^{-1}
\nonumber \\
& &
(\partial -v_{n+2}-w_{n+2})(\partial -u_{n+1}-v_{n+1})
(\partial -u_{n}-w_{n+1}).
\label{llax}
\end{eqnarray}
We constraint this system 
under the two-periodic condition 
\begin{equation}
\Psi_{n+2}=\mu\Psi_{n},\;\;\;
\tilde{\Psi}_{n+2+\frac{1}{3}}=\mu
\tilde{\Psi}_{n+\frac{1}{3}},\;\;\;
\hat{\Psi}_{n+2+\frac{2}{3}}=\mu
\hat{\Psi}_{n+\frac{2}{3}},
\end{equation}
where $\mu$ is an arbitrary constant.
{}Furthermore 
we set the total energy,
\begin{equation}
{\cal H}_{1}=-\sum_{n}(u_{n}+v_{n}+w_{n})=0.
\label{energy}
\end{equation}
Using (\ref{llax}) we can obtain 
\begin{eqnarray}
\lambda\mu\Psi_{n}&=&L_{n}^{(1)}\Psi_{n},
\;\;\;\mu\Psi_{n}=L_{n}^{(2)}\Psi_{n},
\nonumber \\
\lambda\mu\tilde{\Psi}_{n+\frac{1}{3}}&=&
\tilde{L}_{n+\frac{1}{3}}^{(1)}\tilde{\Psi}_{n+\frac{1}{3}},
\;\;\;
\mu\tilde{\Psi}_{n+\frac{1}{3}}=
\tilde{L}_{n+\frac{1}{3}}^{(2)}\tilde{\Psi}_{n+\frac{1}{3}},
\nonumber \\
\lambda\mu\hat{\Psi}_{n+\frac{2}{3}}&=&
\hat{L}_{n+\frac{2}{3}}^{(1)}\hat{\Psi}_{n+\frac{2}{3}},
\;\;\;
\mu\hat{\Psi}_{n+\frac{2}{3}}=
\hat{L}_{n+\frac{2}{3}}^{(2)}\hat{\Psi}_{n+\frac{2}{3}}.
\end{eqnarray}
where 
\begin{eqnarray}
L_{n}^{(2)}&=&(\partial -u_{n}-v_{n}-w_{n})
(\partial -u_{n+1}-v_{n+1}-w_{n+1}),
\nonumber \\
L_{n}^{(1)}&=&
(\partial -u_{n+1}-v_{n+1})(\partial -u_{n}-w_{n+1})
(\partial -v_{n}-w_{n}),
\nonumber \\
\tilde{L}_{n+\frac{1}{3}}^{(2)}&=&
(\partial -u_{n}-v_{n}-w_{n+1})
(\partial -u_{n+1}-v_{n+1}-w_{n}),
\nonumber \\
\tilde{L}_{n+\frac{1}{3}}^{(1)}&=& 
(\partial -u_{n+1}-w_{n})(\partial -v_{n+1}-w_{n+1})
(\partial -u_{n}-v_{n}),
\nonumber \\
\hat{L}_{n+\frac{2}{3}}^{(2)}&=&
(\partial -u_{n}-v_{n+1}-w_{n+1})
(\partial -u_{n+1}-v_{n}-w_{n}),
\nonumber \\
\hat{L}_{n+\frac{2}{3}}^{(1)}&=&
(\partial -v_{n}-w_{n})(\partial -u_{n+1}-v_{n+1})
(\partial -u_{n}-w_{n+1}).
\label{llax2}
\end{eqnarray}
Then the Lax operators split into two parts.
The two parts are 
the 3 and 2 reduced type respectively for  the constraint (\ref{energy}). 
It is a torus.
But the equations of motion are still 
3-hungry Volterra equations.

{}Furthermore
we set 
\begin{equation}
\Psi_{n}=\Psi_{n+1}=\Psi,
\;\;\;
\tilde{\Psi}_{n+\frac{1}{3}}
=\tilde{\Psi}_{n+\frac{4}{3}}=\tilde{\Psi},\;\;\;
\hat{\Psi}_{n+\frac{2}{3}}
=\hat{\Psi}_{n+\frac{5}{3}}=\hat{\Psi},
\end{equation}
then the eigenvalue problems become
\begin{equation}
\lambda\Psi=L^{(1)}\Psi,
\;\;\;
\lambda\tilde{\Psi}=\tilde{L}^{(1)}\tilde{\Psi},
\;\;\;
\lambda\hat{\Psi}=\hat{L}^{(1)}\hat{\Psi},
\end{equation}
where
\begin{eqnarray}
L^{(1)}&=&
(\partial -u-v)(\partial -u-w)
(\partial -v-w),
\nonumber \\
\tilde{L}^{(1)}&=& 
(\partial -u-w)(\partial -v-w)
(\partial -u-v),
\nonumber \\
\hat{L}^{(1)}&=&
(\partial -v-w)(\partial -u-v)
(\partial -u-w).
\label{w3}
\end{eqnarray}
{}From (\ref{w3}) 
we can obtain the classical $W_{3}$ algebra.

On the other hand 
if we set
\begin{equation}
\tilde{\Psi}=\hat{\Psi}=0,
\end{equation}
the spectral problem becomes
\begin{equation}
\lambda\mu\Psi_{n}
=
L_{n}^{(2)}\Psi_{n},
\end{equation}
where
\begin{equation}
L_{n}^{(2)}=(\partial -u_{n})
(\partial -u_{n+1})
\label{vira}
\end{equation}
(\ref{vira}) is nothing but the classical 
 Virasoro algebra.

\section{Two  Matrix Model and Hungry Volterra equation}
\subsection{Hungry Volterra equation}

We consider the computation of the following integral over two Hermitian 
matrices $U$ and $V$ of the size $N\times N$ (see \cite{mm} for a review)
\begin{equation}
Z(t_{3k},\tilde{t}_{3k},N)=\int{\rm d}U{\rm d}V
e^{-{\rm Tr}V(t,\tilde{t})},
\end{equation}
with the potential 
\begin{equation}
V(U,V,t_{3k},\tilde{t}_{3k})
=
\sum_{k=1}^{m_{1}}t_{3k}U^{3k}+\sum_{k=1}^{m_{2}}\tilde{t}_{3k}V^{3k}
+gUV,
\label{pot}
\end{equation}
and with respect to the Haar measure over Hermitian matrices $H$
\begin{equation}
{\rm d}H=c_{N}\prod _{i=1}^{N}{\rm d}H_{ii}\prod_{1\leq i<j\leq N}
{\rm d}{\rm Re}H_{ij}{\rm d}{\rm Im}H_{ij}.
\end{equation}
The parameters $t_{3k},\tilde{t}_{3k},g$ are  real numbers.
We reduce the integral to an integral over the eigenvalues 
of $U$ and $V$, denoted by $u_{i}$ and $v_{i}$ respectively.
In the well known method we can obtain
the reduced integral
\begin{equation}
Z(t,\tilde{t}_{3k};N)
=
\int{\rm d}u{\rm d}v
\Delta(u)\Delta(v)e^{-{\rm Tr}V(u,v,t_{3k},\tilde{t}_{3k})},
\label{mi}
\end{equation}
where 
$\Delta(u)=\prod_{i<j}(u_{i}-u_{j})$ and $\Delta(v)=\prod_{i<j}(v_{i}-v_{j})$
  denote  the Vandermonde determinant
of matrix $u$ and $v$
where 
$u={\em diag}(u_{1},\cdots,u_{N})$ and  $v={\em diag}(v_{1},\cdots,v_{N})$.
Here even if we rotate  the eigen values 
\begin{equation}
u_{i}\rightarrow e^{{\rm i}\frac{2\pi}{3}}u_{i},
\;\;\;\;
v_{i}\rightarrow e^{-{\rm i}\frac{2\pi}{3}}v_{i},
\label{inv}
\end{equation}
(\ref{mi}) does not change 
from the form of the potential (\ref{pot}).
As in the standard orthogonal polynomial technique, 
we introduce two sets of polynomials $p_{n}=x^{n}+$lower degree,
 and $\tilde{p}_{n}(y)=y^{n}+$lower degree , for $n=0,1,2,\cdots$,  
which are orthogonal with respect to the one dimensional 
measure inherited from (\ref{mi}),
namely
\begin{equation}
(p_{n},\tilde{p}_{m})=\int
{\rm d}x{\rm d}ye^{-V(t_{3k},\tilde{t}_{3k})}
p_{n}(x)\tilde{p}_{m}(y)=h_{n}\delta_{m,n}.
\end{equation}
Using the multi-linearity of the determinants, we may rewrite 
\begin{eqnarray}
\Delta(u)&=&{\rm det}[u_{i}^{j-1}]_{1\leq i,j\leq N}=
{\rm det}[p_{j-1}(u_{i})]_{1\leq i,j\leq N},
\nonumber \\
\Delta(v)&=&{\rm det}[v_{i}^{j-1}]_{1\leq i,j\leq N}=
{\rm det}[\tilde{p}_{j-1}(v_{i})]_{1\leq i,j\leq N}.
\end{eqnarray}
Using the orthogonality relations, the partition function  is finally can 
be written as 
\begin{equation}
Z(t_{3k},\tilde{t}_{3k};N)
={\rm const.}\prod_{i=0}^{N-1}h_{i}.
\end{equation}
Using  the fact (\ref{inv})
 we can obtain a  relation 
\begin{equation}
p_{n}(x e^{\frac{2\pi{\rm i}}{3}})
\tilde{p}_{n}(y e^{-\frac{2\pi{\rm i}}{3}})
=
p_{n}(x)p_{n}(y).
\label{con}
\end{equation}
The multiplication by $x$ and $y$ can be represented 
by the Jacobi 
matrixes  $Q$ and $\tilde{Q}$:
\begin{equation}
xp_{n}(x)=\sum_{m=0}^{n+1}Q_{n,l}p_{m}(x),\;\;\;
yp_{n}(y)=\sum_{m=0}^{n+1}\tilde{p}_{m}(y)\tilde{Q}_{m,n},
\label{rr}
\end{equation}
where $Q_{nl}$ and $\tilde{Q}_{nl}$ are the matrix elements 
of $Q$ and $\tilde{Q}$.

{}From the definition
of orthogonal polynomials it follows that 
\begin{equation}
Q_{n,n+1}=1,\;\;\;
Q_{n,m}=0,\;\;\;m\geq n+2,\;\;\;\;\;\;
\tilde{Q}_{n+1,n}=1,\;\;\;
\tilde{Q}_{m,n}=0,\;\;\;m\geq n+2,
\end{equation}
Define a wave function
\begin{equation}
\Phi_{n}(t_{3k},\tilde{t}_{3k},x)=
p_{n}(x)e^{V(t_{3k},\tilde{t}_{3k},x)}.
\label{wf}
\end{equation}
Here we introduce the matrix $\bar{Q}$
\begin{equation}
\bar{Q}_{n,m}=
(H\tilde{Q}H^{-1})_{n,m},\;\;\;
H_{n,m}=h_{n}\delta_{nm}.
\end{equation}

In what follows it will be convenient to define an explicit 
parameterization
of matrices $Q$ and $\bar{Q}$.
We choose the following parameterization 
\begin{eqnarray}
Q_{n,n+1}&=&1,\;\;\; Q_{n,n-k}=f_{k}(n),\;\;\; k=0,1,\cdots,3m_{2}-1,
\nonumber \\
Q_{n,m}&=&0,\;\;\; {\rm for }\;\;\;m-n\geq 2,\;\;\;{\rm and}
\;\;\;n-m\geq 3m_{2}-1,
\end{eqnarray}
\begin{eqnarray}
\bar{Q}_{n,n-1}&=&R_{n},\;\;\; \bar{Q}_{n,n+k}=g_{k}(n)
R_{n+1}^{-1}\cdots R_{n+k}^{-1},\;\;\; k=0,1,\cdots,3m_{1}-1,
\nonumber \\
Q_{n,m}&=&0,\;\;\; {\rm for }\;\;\;n-m\geq 2,\;\;\;
{\rm and}\;\;\;m-n\geq 3m_{2}-1,
\end{eqnarray}
where $R_{n}=h_{n}/h_{n-1}$.
To satisfy the condition (\ref{con}) 
we can obtain 
\begin{equation}
f_{3j}(n)=0,\;\;\;
f_{3j+1}(n)=0,\;\;\;
g_{3j}(n)=0,\;\;\;
g_{3j+1}(n)=0,\;\;\;
j=0,1,2,\cdots.
\end{equation}
We can rewrite (\ref{rr})
\bph
\begin{equation}
xp_{n}=p_{n+1}+\sum_{j=1}^{m_{2}}f_{3j-1}(n)p_{n+1-3j},
\label{EP}
\end{equation}
\begin{equation}
y\tilde{p}_{n}=\frac{h_{n}}{h_{n-1}}\tilde{p}_{n-1}+
\sum_{j=1}^{m_{1}}g_{3j-1}(n)
\frac{h_{n}}{h_{n-1+3j}}
\tilde{p}_{n-1+3j}.
\label{EP1}
\end{equation}
\eph
{}From (\ref{wf}) we can obtain
\begin{equation}
x\Phi_{n}=Q\Phi_{n},
\;\;\;
\frac{\partial }{\partial t_{r}}\Phi_{n}
=Q^{r}_{+}\Phi_{n},
\;\;\;
\frac{\partial }{\partial \tilde{t}_{s}}\Phi_{n}
=-Q^{s}_{-}\Phi_{n},
\end{equation}
where 
the subscripts  ``+''  denotes 
upper triangular plus diagonal parts of the matrix and 
``-''  denotes lower triangular part.

{}For the first one we can obtain
\begin{eqnarray}
\frac{\partial }{\partial t_{3}}\Phi_{n}
&=&
\Phi_{n+3}
+s_{0}(n)
\Phi_{n},
\nonumber \\
\frac{\partial }{\partial \tilde{t}_{3}}\Phi_{n}
&=&
-\frac{h_{n}}{h_{n-3}}\Phi_{n-3}
,
\label{ii}
\end{eqnarray}
where 
\begin{equation}
s_{0}(n)=f_{2}(n)+f_{2}(n+1)+f_{2}(n+2). 
\end{equation}
Hereafter we  consider 
the special case $m_{2}=1$.
In this case the recursion relation  (\ref{EP}) becomes 
\begin{equation}
x^{3}\Phi_{n}
=
\Phi_{n+3}+s_{0}(n)\Phi_{n}
+s_{1}(n)\Phi_{n-3}+s_{2}(n)\Phi_{n-6},
\label{ep}
\end{equation}
where
\begin{equation}
s_{1}(n)=f_{2}(n+1)f_{2}(n-1)
+f_{2}(n-1)f_{2}(n)+f_{2}(n)f_{2}(n-2),\;\;\;
s_{2}(n)=
f_{2}(n)f_{2}(n-2)f_{2}(n-4).
\end{equation}
(\ref{ii}) and (\ref{ep}) 
are nothing but the eigenvalue  problem (\ref{flow}) and (\ref{see}).
Then consistency of (\ref{ii}) and (\ref{ep}) becomes 
the hungry Vol-Terra equation
\begin{equation}
\frac{\partial f_{2}(n)}{\partial t_{3}}
=
f_{2}(n)
(f_{2}(n+2)+f_{2}(n+1)-f_{2}(n-1)-f_{2}(n-2)).
\label{hvv}
\end{equation}
To the flows $t_{3k}$ we can obtain 
the higher order hierarchy of (\ref{hvv}).

On the other hand
from (\ref{ii}) and (\ref{ep})
\begin{equation}
\frac{\partial {\cal R}_{n}}
{\partial t_{3}}=
{\cal R}_{n}(s_{0}(n)-s_{0}(n-3)),\;\;\;
\frac{\partial s_{0}(n)}{\partial \tilde{t}_{3}}
={\cal R}_{n+3}-{\cal R}_{n},
\label{2toda}
\end{equation}
where 
${\cal R}_{n}=h_{n}/h_{n-3}.$
(\ref{2toda}) is nothing but the 2-dimensional 
Toda equation.

We  consider  the most simple case $m_{1}=m_{2}=1$. We set $g=N$ and $t_{3}
=\tilde{t}_{3}=-tN$.
If we think of $t$ as a small parameter,
we may at least formally expand the integral $Z$ in power of $t$.
This expansion is expressible as a sum over Feynmann graphs,
 made of double lines oriented in  opposite  directions and carrying a matrix index
$i\in \{1,2,\cdots,N\}$, involving two types of  three-valent vertices,  connected 
by one type of propagator. ($<UV>$ propagator, weighted by $1/N$)
As  usual  in matrix integrals, each graph receives a contribution 
$N$ per oriented loops and $tN$ per vertices.
 We finally obtain   the expansion
\begin{equation}
F(t,N)=\log Z(t,N)=
\sum_{\Gamma}\frac{t^{V(\Gamma)}N^{\chi(\Gamma)}}{|{\rm Aut}(\Gamma)|},
\label{15}
\end{equation}
where  
$V(\Gamma)$ denotes the number of 
edges, 
$\chi(\Gamma)$ its Euler characteristic,
$|{\rm Aut}(\Gamma)|$ denotes the order of the automorphism group of 
bi-colored connected graphs  $\Gamma$.
The bi-colored  graphs mean
that two adjacent vertices have different colors.
We can consider ({\ref{15}) as an expansion over dual graphs, 
with now black and white alternating faces.
Then  we can finally view $\log Z$ as the partition function 
for  black and white colored triangurations.

\subsection{Discrete time hungry Volterra equation}

In this subsection we consider following  two matrix model 
\begin{equation}
Z(t_{3k},\tilde{t}_{3},l)=\int{\rm d}U{\rm d}V
e^{-{\rm Tr}V(t,\tilde{t},l)}
\label{++}
\end{equation}
with the potential 
\begin{equation}
V(U,V,t_{3k},\tilde{t}_{3},l)
=
\sum_{k=1}^{m_{1}}t_{3k}U^{3k}+\tilde{t}_{3}V^{3}
+gUV-3l\log U
\label{pot2}
\end{equation}
We added  the ``log type potential''.\cite{pe}\cite{hm}
Note that $l=0,1,2,\cdots$.
In the special case $l=0$,
 (\ref{++}) becomes the matrix model studied 
in the previous subsection.
The orthogonal polynomials satisfy the orthogonal relations
\begin{equation}
(p_{n,l},\tilde{p}_{m,l})=\int
{\rm d}x{\rm d}ye^{-V(t_{3k},\tilde{t}_{3k},l)}
p_{n,l}(x)\tilde{p}_{m,l}(y)=h_{n,l}\delta_{m,n}.
\label{relation}
\end{equation}
The partition function (tau function)
can be written 
\begin{equation}
Z(t_{3},\tilde{t}_{3k},l;N)
={\rm const.}\prod_{i=0}^{N-1}h_{i,l}.
\label{pf}
\end{equation}
We can obtain the recursion relations 
of orthogonal polynomials in the same way 
\bph
\begin{equation}
xp_{n,l}=p_{n+1,l}+f_{2}(n,l)p_{n-2,l},
\label{lp1}
\end{equation}
\begin{equation}
y\tilde{p}_{n,l}=\frac{h_{n,l}}{h_{n-1,l}}\tilde{p}_{n-1,l}+
\sum_{j=1}^{m_{1}}g_{3j-1}(n,l)
\frac{h_{n,l}}{h_{n-1+3j.l}}
\tilde{p}_{n-1+3j,l}.
\label{EP2}
\end{equation}
\eph
Notice that orthogonal polynomials $p_{n,l}$ and 
$\tilde{p}_{n,l}$ and  $f(n,l)$, $g(n,l)$ and $h_{n,l}$
are  the functions of $n$ and $l$.
We rewrite $f_{2}$ using the partition function (\ref{pf})
\begin{equation}
f_{2}(n,l)\equiv V_{n,l}=\frac{Z_{n+3,l}Z_{n-2,l}}{Z_{n+1,l}Z_{n,l}}.
\end{equation}
Using (\ref{relation}) and the condition 
$e^{-V(t_{3k},\tilde{t}_{3k},l+1)}=x^{3}e^{-V(t_{3k},\tilde{t}_{3k},l)}$
 we can get a  relation
\begin{equation}
x^{3}p_{n,l+1}=p_{n+3,l}+I_{n,l}p_{n,l},
\;\;\;
I_{n,l}=\frac{h_{n,l+1}}{h_{n,l}}=\frac{Z_{n+1,l+1}Z_{n-2,l}}{Z_{n-2,l+1}
Z_{n+1,l}}.
\label{tf}
\end{equation}
As a compatibility condition of 
(\ref{lp1}) and (\ref{tf})   we can obtain 
\begin{equation}
I_{n+1,l}+V_{n,l+1}=V_{n+3,l}+I_{n,l},\;\;\;
I_{n,l}V_{n,l}=V_{n,l+1}I_{n-2,l}.
\label{emk}
\end{equation}
Here we change dependent variables $V_{(n,l)}\rightarrow \tilde{V}_{(n,l)}$.
Notice that $V_{n,l}=f_{2}(n,l)$ are  the dependent variables 
of the hungry Volterra equation (\ref{hvv}).
We define
\begin{equation}
V_{n,l}=
\tilde{V}_{n,l}(1+\tilde{V}_{n,l-1})(1+\tilde{V}_{n,l-2}).
\end{equation}
Using $\tilde{V}_{n,l}$ 
we can rewrite
\begin{equation}
I_{n.l}=(1+\tilde{V}_{n,l})(1+\tilde{V}_{n,l-1})(1+\tilde{V}_{n,l-2}).
\end{equation}
The equation of  motion 
(\ref{emk}) becomes
\begin{equation}
\frac{\tilde{V}_{n,l+1}}{\tilde{V}_{n,l}}
=\frac{(1+\tilde{V}_{n-1,l})(1+\tilde{V}_{n-2,l})}
{(1+\tilde{V}_{n-1,l+1})(1+\tilde{V}_{n-2,l+1})}.
\label{dhv}
\end{equation}
(\ref{dhv}) is the discrete hungry Voltrra equation.
\cite{pn}-\cite{s5}
Using the partition function (\ref{pf}) we  can  rewrite 
\begin{equation}
\tilde{V}_{n,l}=
\frac{Z_{n+3,l}Z_{n-2,l+1}}{Z_{n,l+1}Z_{n+1,l}},
\;\;\;
1+\tilde{V}_{n,l}
=
\frac{Z_{n,l}Z_{n+1,l+1}}{Z_{n+1,l}Z_{n,l+1}}.
\end{equation}
(\ref{dhv}) is  cast into the following bilinear equation
\begin{equation}
Z_{n-2,l+1}Z_{n+3,l}+Z_{n+1,l}Z_{n,l+1}=Z_{n,l}Z_{n+1,l+1}.
\end{equation}

\section{Concluding Remarks}

In this paper we consider the $k$-hungry Volterra equation 
as B\"{a}cklund transformations (BT) of the multi boson  KP equation 
which is related 
to the $sl(k+1, k)$ algebra.
We  call the BT 
``fractional BT''.
If we repeat  the fractional BT  $k$ times 
we can obtain the usual BT which is the Toda lattice.

To study the relations 
between the 2d Toda lattice and the hungry Volterra equation
we consider  the hungry Volterra equation
in the two matrix model.
If we select the time $t_{3k}$ and $\tilde{t}_{3}$ 
from  the flows  of 2d Toda lattice  $t=(t_{1},t_{2},\cdots)$ and 
$\tilde{t}=(\tilde{t}_{1},\tilde{t}_{2},\cdots)$, 
we can obtain the 3-hungry Volterra hierarchy  which 
is related to the   lattice $W_{3}$.
The  most simple case is the matrix model for the bi-colored random triangulation.
It is easy to  construct  the general $k$-hungry Volterra equation in 
2d Toda lattice.
Using the duality of  $t$ and $\tilde{t}$ 
we can obtain two $k$-hungry Volterra hierarchies in 2d Toda lattice.

{}Furthermore we 
consider the discrete time hungry Volterra equation
in the two matrix model with 
the Penner type potential.
It is well known that 
the partition function of the one (multi) matrix model with 
the Penner type potential  satisfies  the 
discrete time Toda equation.\cite{hm}\cite{df}\cite{ka}
We select the flows $l$ and $\tilde{t}_{3}$ where 
$l$ is the multiple of 3 of discrete time lattice   which  belongs to  the 
2d  discrete time Toda equation.
If we select the multiple of $k$  of discrete time lattice 
and $\tilde{t}_{k}$,  we can obtain the 
discrete time $k$-hungry Volterra equation.
Then we can find two discrete time  $k$-hungry Volterra equations 
in  2d discrete time Toda equation.

\end{document}